\documentclass{iaus}
\usepackage{graphicx}

\newcommand\ga{\gtrsim}
\newcommand\la{\lesssim}
\newcommand\micron {\, \mu \hbox{m}}

\newcommand\percc{\mathrm{cm}^{-3}}

\def\ee #1 {\times 10^{#1}}          
\def\ut #1 #2 { \, \textrm{#1}^{#2}} 
\def\un #1 { \, \textrm{#1}}          
\def\u #1 { \, \textrm{#1}}          
\def\kms {\,\textrm{km\,s}^{-1}}

\title[OH Masers and Supernova Remnants] 
{OH Masers and Supernova Remnants}

\author[Mark Wardle \& Korinne McDonnell]   
{Mark Wardle 
 \and Korinne McDonnell}

\affiliation{Department of Physics \& Astronomy and Research Centre for Astronomy, Astrophysics \& Astrophotonics, Macquarie University, Sydney NSW 2109, Australia \\ email: {\tt mark.wardle@mq.edu.au, korinne.mcdonnell@mq.edu.au} }

\pubyear{2012}
\volume{287}  
\pagerange{1--2}
\jname{Cosmic Masers -- from OH to H$_0$}
\editors{R. Booth, E. Humphries \& W. Vlemmings, eds.}
\begin{document}

\maketitle

\begin{abstract}
OH(1720 MHz) masers are created by the interaction of supernova remnants with molecular clouds.  These masers are pumped by collisions in warm, shocked molecular gas with OH column densities in the range $10^{16}$--$10^{17}$\,cm$^{-2}$.   Excitation calculations suggest that inversion of the 6049\,MHz OH line may occur at the higher column densities that have been  inferred from main-line absorption studies of supernova remnants with the Green Bank Telescope.  OH(6049\,MHz) masers have therefore been proposed as a complementary indicator of remnant-cloud interaction.

This motivated searches for 6049\,MHz maser emission from supernova remnants using the Parkes 63\,m and Effelsberg 100\,m telescopes, and the Australia Telescope Compact Array.  A total of forty-one remnants have been examined by one or more of these surveys, but without success.    To check the accuracy of the OH column densities inferred from the single-dish observations we modelled OH absorption at 1667\,MHz observed with the Very Large Array towards three supernova remnants, IC 443, W44 and 3C 391.  The results are mixed -- the OH column is revised upwards in IC443, downwards in 3C391, and is somewhat reduced in W44.  We conclude that OH columns exceeding $10^{17}$\,cm$^{-2}$ are indeed present in some supernova remnants and so the lack of any detections is not explained by low OH column density.   We discuss the possibility that non-local line overlap is responsible for suppressing the inversion of the 6049 MHz line.

\end{abstract}

\firstsection 
\section{Introduction}

OH (1720\,MHz) masers associated with supernova remnants are unambiguous indicators of interaction with adjacent molecular clouds (Frail, Goss \& Slysh 1994). The relative ease of identifying supernova remnant - molecular cloud interaction sites relative to other methods, such as searching for molecular gas that is kinematically disturbed or chemically processed (e.g. DeNoyer 1979a,b; Wootten 1981) motivated surveys of the galaxy for more examples (Frail et al 1996; Yusef-Zadeh et al 1996; Green et al 1997; Koralesky et al 1998; Yusef-Zadeh et al 1999a,b; Sjouwerman \& Pihlstr\"om 2008), which have found that approximately 10\% of galactic supernova remnants have 1720\,MHz masers and are therefore interacting with molecular clouds.    

\begin{figure}
\begin{center}
      \includegraphics[trim=0 0 30 0, clip=true, scale=0.5]{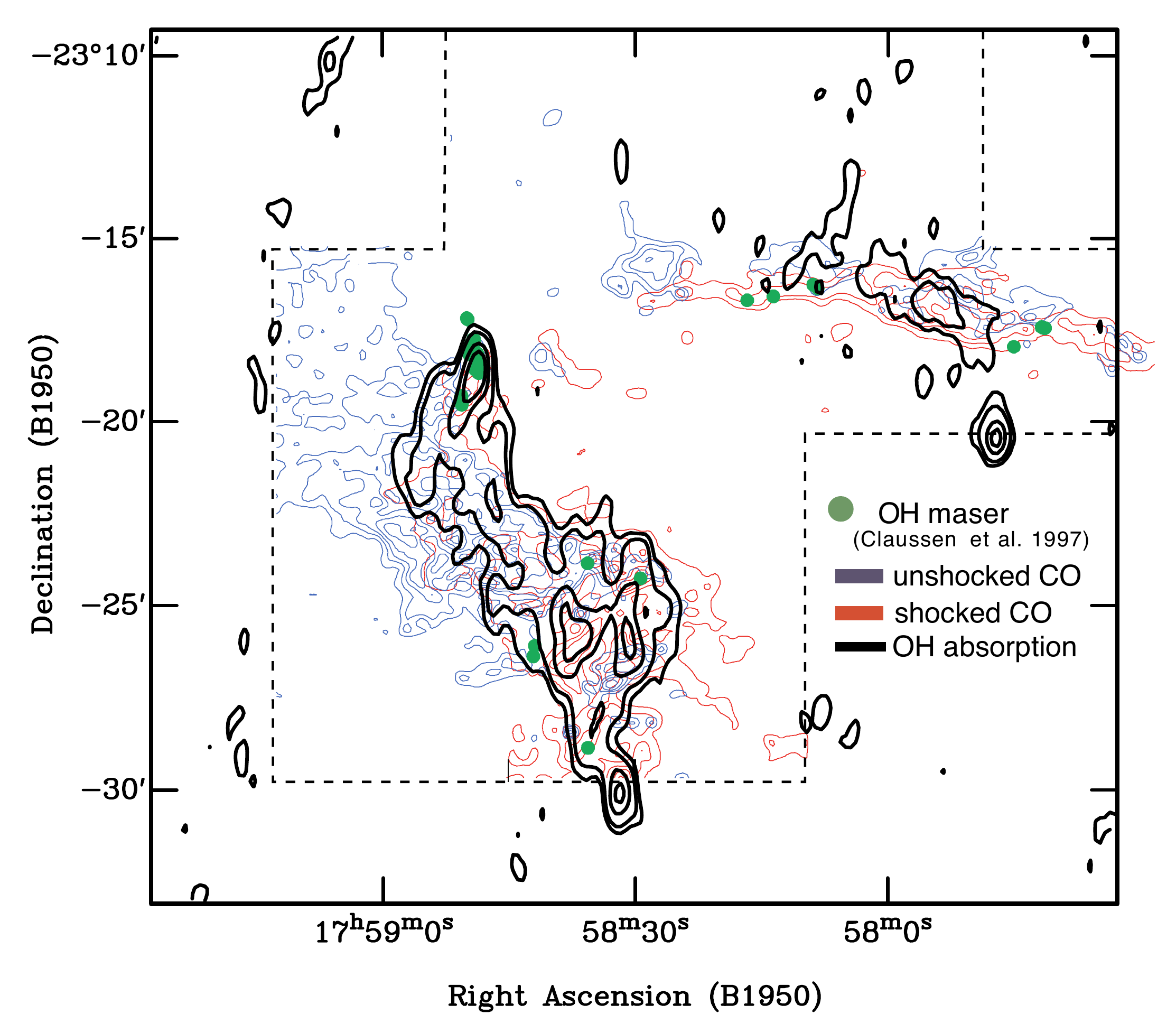}
\caption{\label{fig:W28}OH absorption from ATCA observations of SNR W28 (Green, Wardle \& Lazendic 2000),  overlaid on the locations of bright OH(1720 MHz) maser spots (Claussen et al.~1997), and contours of CO J=1--0 and J=3--2 emission (Arikawa et al.~1999) tracing unshocked and shocked gas, respectively. (Adapted from Arikawa et al.~1999)}
\end{center}
\end{figure}

The masers are collisionally pumped in warm, dense molecular gas with densities and temperatures in the range 50--125\,K, $\sim 10^3$--$10^5\percc$ and OH column densities $10^{16}$--$10^{17}\ut cm -2 $ (Elitzur 1976; Lockett, Gauthier \& Elitzur 1999).  These conditions are not met in gas overrun by standard ``J-type'' shock waves which dissociate molecules:  although molecules reform as the shocked gas cools (e.g.~ Hollenbach \& McKee 1989), cooling is so rapid that the OH column that is produced in the relevant temperature range is too small (Lockett, Gauthier \& Elitzur 1999). Instead the molecular shock must be ``C-type''(Mullan 1971; Draine 1980), in which heating and compression of the gas within the shock front is achieved by magnetically-driven ion-neutral collisions.  The low fractional ionisation of the molecular gas means that this process is relatively slow, yielding a broad shock front with an extended warm tail.  However the requisite OH column is still higher than predicted by standard C-type shock models which rapidly convert OH into water (Draine, Roberge \& Dalgarno 1983; Kaufman \& Neufeld 1996). Instead an additional agent, such as X-rays or cosmic rays associated with the supernova remnant, is needed to dissociate water into OH as the gas cools behind the shock,   (Wardle 1999).

The clear association of 1720\,MHz masers with shocked gas (see e.g.~Fig. \ref{fig:W28}), underpinned by a theoretical production mechanism hinging on molecular shock waves, provides a firm basis for assuming that the presence of these masers signals the existence of a shock wave driven by a supernova remnant into an adjacent cloud (Wardle \& Yusef-Zadeh 2002).  This has been confirmed by follow-up studies of individual remnants.  A good illustration of this is provided by the supernova remnant G349.7+0.2, about which little was known apart from its having a radio continuum shell and four OH(1720 MHz) masers.    A search for CO (J=1--0) emission from an adjacent molecular cloud,  H$_2$ (v=1--0) emission at $2.12\,\micron$ from shocked molecular gas, and OH absorption associated with the masers found all three (Lazendic et al.~2010).  The population of 1720\,MHz-bearing supernova remnants show other hallmarks of interaction.  For example ``mixed-morphology'' remnants are characterised a radio shell, or partial shell, filled with centrally-peaked emission in soft thermal x-rays (Long et al.~1991; Rho \& Petre 1998) believed to be provided by the evaporation of dense interstellar gas in the interior of the remnants (White \& Long 1991; Cox et al.~1999).  There is a strong statistical association between remnants hosting 1720 MHz masers and mixed-morphology remnants (Yusef-Zadeh et al.~2003), and also with those with associated gamma-ray sources (Hewitt et al.~2009), which are likely produced by interactions of cosmic-rays accelerated in the remnant with the neighbouring molecular gas.

Apart from signalling locations where remnants are driving shock waves into molecular clouds, 1720\,MHz masers are important probes of the interaction. The masers require significant columns of shocked molecular gas and good velocity coherence, and so tend to occur on the limbs of supernova remnant shells at velocities close to the systemic LSR velocity of the remnant, allowing distance estimates based on models of galactocentric rotation.  Follow-up Zeeman observations at 1720 MHz have determined that the magnetic field in the post-shock gas is $\sim1$--3\,mG (Claussen et al 1997, 1999; Brogan et al 2000; Hoffman et al 2005).  The post-shock magnetic field acts as the piston driving a C-type shock wave, and a field of this strength provides sufficient post-shock pressure needed to drive a shock at 25$\kms$ into gas of density $\sim10^{4}\,\percc$, so these measurements provide an important sanity check on the basic picture.   In addition the magnetic field measurements provides a good estimate of the pressure in the interior of the supernova remnant that is driving the shock into the molecular cloud.

\section{Searches for 6049\,MHz masers}
While 1720 MHz masers are invaluable in pinpointing where supernova remnants interact with molecular clouds, their absence does not imply that an interaction is \emph{not} occurring.  Instead,  one or more of the physical conditions required to collisionally invert the 1720\, MHz transition may simply be absent  -- a sufficient column of OH, or line-of-sight velocity coherence within the OH column.  Additional indicators of such interactions would therefore be useful.

The 1720 MHz inversion dies away for OH columns $\ga 10^{17}\ut cm -2 $ due to photon trapping, but under similar physical conditions the analogous 6049\,MHz transition in the first rotationally-excited state of OH becomes inverted (Pavlakis \& Kylafis 2000; McDonnell, Wardle \& Vaughan 2008).  
This  prompted searches for masers at 6049\,MHz associated with supernova remnants using Parkes (McDonnell, Wardle \& Vaughan 2008), Effelsberg (Fish, Sjouwerman \& Philstrom 2007) and the ATCA (McDonnell 2011) (see Table 1).  These searches targeted SNR with 1720\,MHz masers, mixed-morphology, or other evidence of interaction (note that there is substantial overlap between the SNR in the Parkes \& Effelsberg and Parkes \& ATCA searches), but all were unsuccessful. 

\begin{table}[htdp]
\caption{Searches for 6049 MHz masers}
\vspace*{6pt}
\begin{center}
\begin{tabular}{@{} lrccl @{}}
Telescope   &  $N_\mathrm{SNR}$  & beam                    &  $\sigma$ (mJy/beam) & Reference   \\
\hline
Parkes      &  35                & $200''$                 &  $\sim 150$ & McDonnell, Wardle \& Vaughan (2008) \\
Effelsberg  &  14                & $130''$                 &  $\sim  10$ & Fish, Sjouwerman \& Philstrom (2007)\\
ATCA        &  17                & $20''\times9''$ &  $\sim  22$ & McDonnell (2011) \\
\hline
\end{tabular}\\
\end{center}
\end{table}

The non-detections can be converted into upper limits on the 6049\,MHz maser optical depth (i.e.~maser amplification factor) using estimates of the brightness temperature of the background continuum and the likely beam filling factor of any putative maser emission.  In turn, an upper limit on the maser optical depth can be translated to an upper limit on OH column density through the OH excitation calculations. It turns out that the relatively poor sensitivity of the Parkes observations does not place strong constraints on the OH column; by contrast, the derived limits on OH column density implied by the lack of detected 6049\,MHz masers in the Effelsberg and ATCA searches are $\la 10^{16.5}\ut cm -2 $ (McDonnell 2011).  These upper limits are at odds with the OH columns exceeding $\sim 10^{17}\ut cm -2 $ that were inferred in nine supernova remnants by modelling the of OH absorption in the 1612, 1665, 1667\,MHz lines observed with the Green Bank Telescope (Hewitt et al.~2006; Hewitt, Yusef-Zadeh \& Wardle 2008).  These nine remnants are listed in Table 2 -- all but G349.7+0.2 were searched for 6049\,MHz emission with Effelsberg and/or ATCA, and so one might have expected several detections. 

One potential source of error is the $7.4$ arc minute beam of the GBT observations.  While a beam filling factor was included in the modelling by Hewitt et al.~(2008) this does not account for different sub-structure in the background continuum and overlying OH absorption within the beam, and given that the inferred beam filling factors are typically $\la 10^{-2}$ this omission may be significant.  To check whether the inferred OH columns  are systematically overestimated we conducted 1667\,MHz OH absorption observations of 3 remnants (3C\,391, W44, and IC\,443) using the VLA in D array, yielding a beam size of $\sim55''\times 90''$.   We followed the approach of Hewitt et al.\ (2008) in modelling the absorption at 1667\,MHz in conjunction with 1720\,MHz VLA data (the latter kindly provided by Jack Hewitt).  

Notably, the derived solid angles subtended by the OH absorption components -- which are typically marginally resolved in the VLA observations -- are in reasonable agreement with the GBT results.   However, the comparison of the derived OH columns gives mixed results: the peak column density derived from the VLA observations for 3C391, W44 and IC443 are much less, a little less, and somewhat greater than the GBT-derived columns, respectively.

\begin{figure}
\begin{center}
      \includegraphics[trim=0 0 0 0, clip=true, scale=0.65]{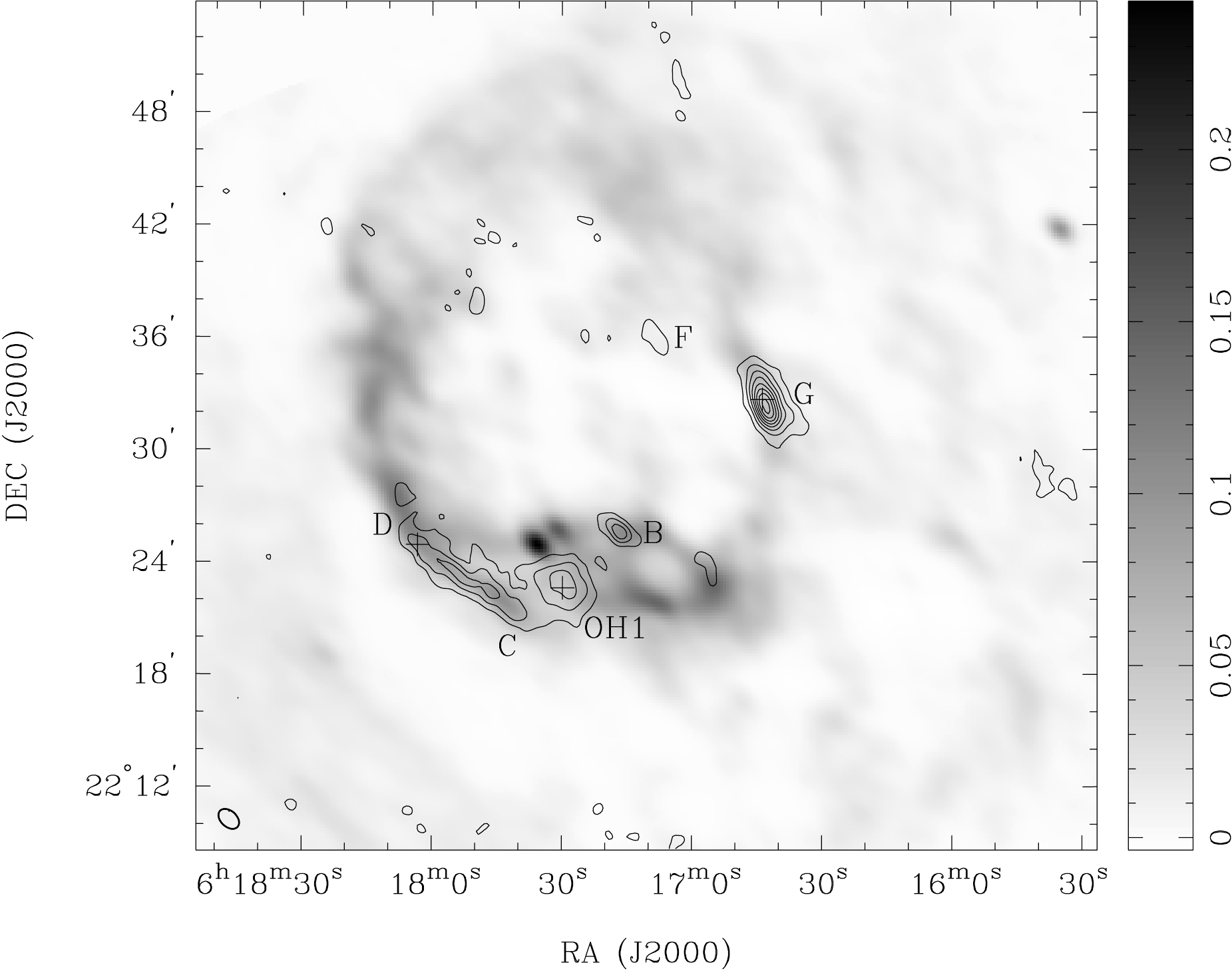}
\caption{ Grey-scale shows the continuum emission of IC 443 at 1667 MHz (Jy per $76''\times 51''$ beam).  The beam size and orientation are indicated by the ellipse in the lower left corner.  Contours show the velocity-integrated OH absorption from 0.2 to 1.6 Jy\,beam$^{-1}$\,km\,s$^{-1}$ in steps of 0.2 Jy\,beam$^{-1}$\,km\,s$^{-1}$.  The three crosses mark positions of 1720 MHz OH maser emission (Hewitt 2009). The OH absorption clumps B, C, D, F, and G are labelled following the identification of shocked CO clumps by Huang, Dickman \& Snell (1986).  The clump of OH absorption labelled OH1 A has not been previously identified.\label{fig:IC443}}
\end{center}
\end{figure}

\begin{figure}
\begin{center}
      \includegraphics[trim=0  0 0 0, clip=true, scale=0.63]{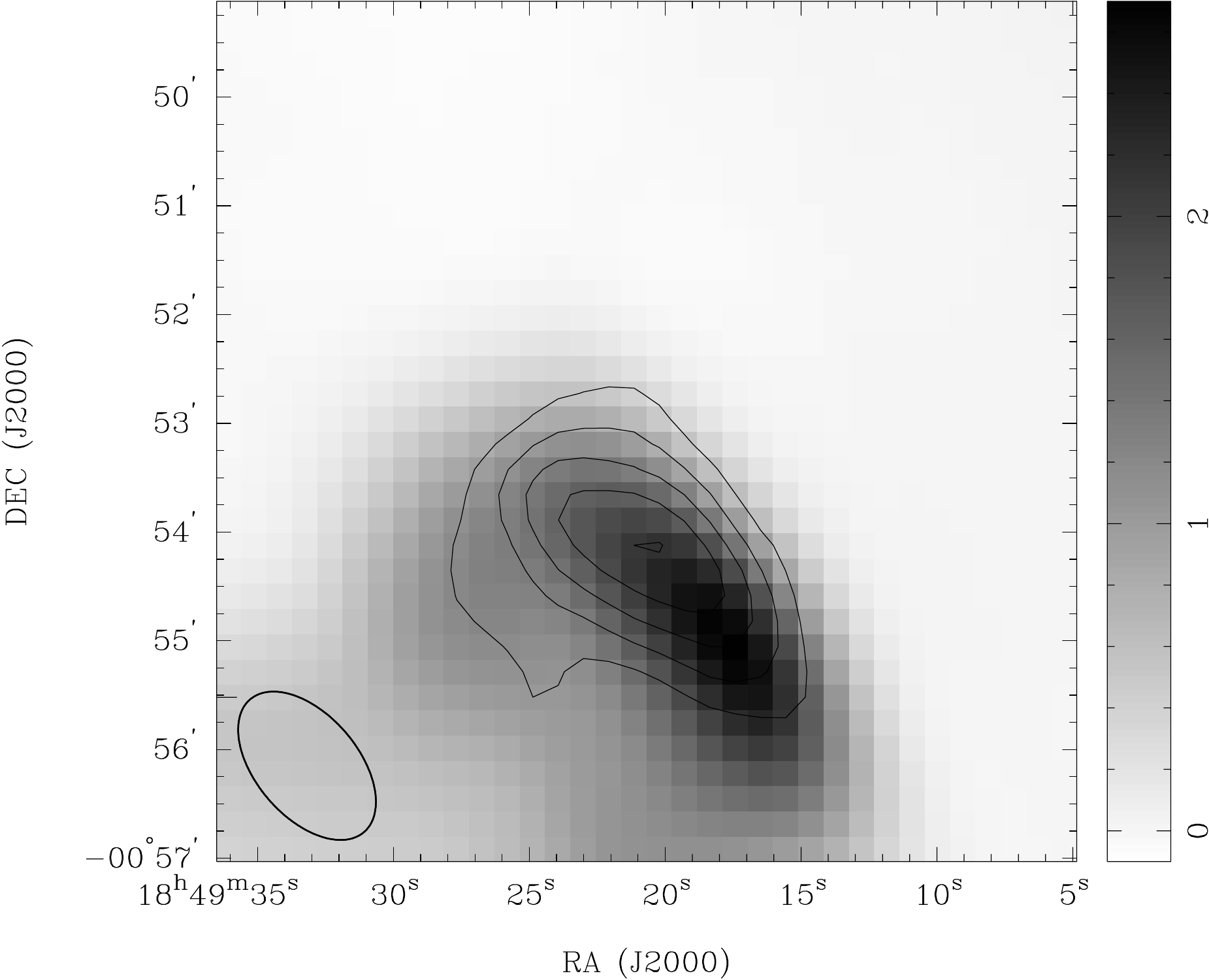}
\caption{Grey-scale shows the continuum emission of the NE corner of 3C\,391 at 1667\,MHz between -0.1 and 2.7\,Jy per $96''\times 56''$ beam.  Absorption in the 1667\,MHz line of OH is indicated by the contours, with levels at $-1$, $-0.8$, $-0.6$, $-0.4$, $-0.2$~Jy~beam$^{-1}$~km~s$^{-1}$.  This image is $\sim$7.44~arcmin on a side, approximately matching the GBT beam area of the 3C~391 spectrum from Hewitt et al.~(2008). \label{fig:3C391}}
\end{center}
\end{figure}

Fig.\ \ref{fig:IC443} shows the 1667\,MHz absorption against the continuum in IC 443.  As expected, OH absorption surrounds the three known 1720\,MHz OH masers, which are associated with molecular clumps clumps D and G and a new feature, labeled OH1.   Focussing on clump G, which has the greatest column density, we derived model parameters quite similar to the GBT.  This reflects that the OH emission is dominated by a single compact component and the background continuum is quite smooth.  For 3C\,491 by contrast, the continuum shows strong structure on the scale of the dominant absorption in the NW of the remnant (see Fig. 3).  In this case the GBT observations, which have a beam area of about the size of the image in Fig. \ref{fig:3C391}, underestimate the strength of the background continuum underlying the absorption by averaging over the beam, resulting in an overestimate of the optical depth in the 1667\,MHz line.   The derived OH column density is therefore too large.


In summary, the VLA observations show that the GBT modelling of the does pretty well, but the poor resolution of single dish observations leads to significant overestimates of the OH column density when the background continuum and OH absorption have poorly correlated substructure within the beam.  Nevertheless IC 443 and potentially some of the other remnants in Table 2 have OH column densities exceeding $10^{17}\ut cm -3 $ and yet were not detected in the ATCA and Effelsberg 6049\,MHz searches.

\begin{table}[htdp]
\caption{SNR with $N_\mathrm{OH} \ga 10^{17} \ut cm -2 $ inferred from GBT observations.\label{Table:SNR-NOH}}
\vspace*{6pt}
\begin{center}
\begin{tabular}{@{} lclc @{}}
SNR         &  $N_\mathrm{OH}$ (dex)   &  6049\,MHz  &  $N_\mathrm{OH}$ (dex)    \\
            &  from GBT                &  searches$^1$   &  from VLA    \\
\hline
W28         &  $17.14 \pm 0.01$ &  P E A              &            \\
G16.7+0.1   &  $16.9  \pm 0.6 $ &  P E A              &            \\
3C391       &  $17.1  \pm 0.4 $ &  P E                &  $ < 16.5$ \\
W44         &  $17.14 \pm 0.01$ &  P E                &  $ < 17.1$ \\
IC443       &  $17.11 \pm 0.01$ &  P E                & $\sim17.5$ \\
CTB37A      &  $17.1  \pm 0.4 $ &  P \quad A                &            \\
G349.7+0.2  &  $17.04 \pm 0.02$ &  P                  &            \\
G357.7-0.1  &  $17.12 \pm 0.01$ &  P \quad A                &            \\
G359.1-0.5  &  $17.13 \pm 0.01$ &  P \quad A                &            \\
\hline
\end{tabular}\\
$^1$ P=Parkes, E=Effelsberg, A=ATCA
\end{center}
\end{table}

\section{Line overlap}
An alternative possibility is that the treatment of overlapping spectral lines in the OH excitation calculations is at fault.  Line overlap in OH occurs even for velocity widths of order $1\kms$ because of the small splitting of the rotational levels arising through lambda doubling and hyperfine splitting.  This enables photons emitted in one infrared transition between sub levels of different rotational states in one OH molecule to to be absorbed by a transition between different sub-levels in the same pair of rotational rates.  This tends to scramble the level populations within the upper or lower pairs of sub levels inside each rotational level and suppress population inversion in the 1720 and 6049\,MHz transitions, quenching the masers.  

\begin{figure}
\begin{center}
      \includegraphics[scale=0.7]{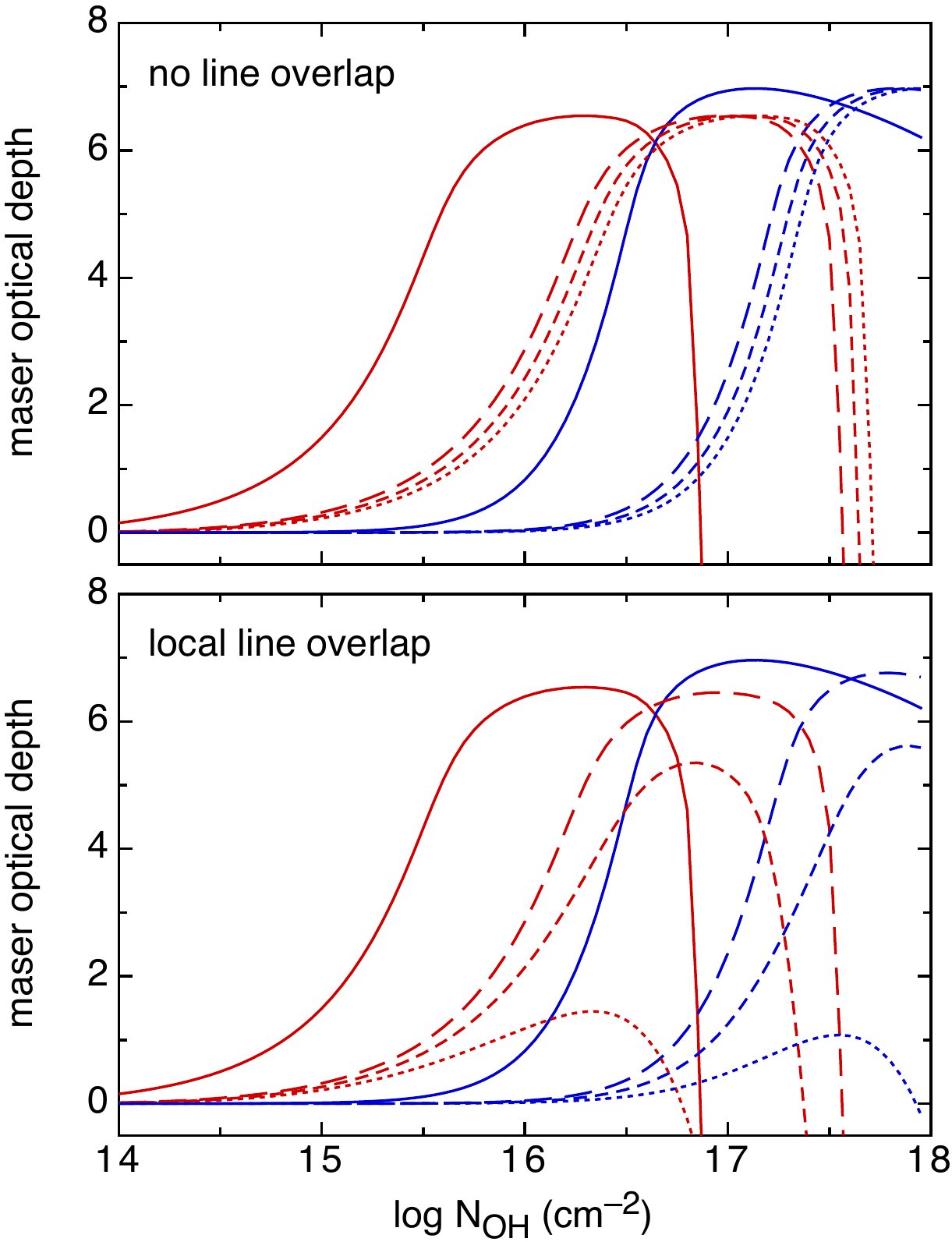}
\caption{1720 MHz (red) and 6049 MHz (blue) maser optical depth as a function of OH column density for a uniform slab with $n(\mathrm{H}_2) = 10^4 \percc$ and $T=50$\,K as the total (thermal + micro-turbulent) FWHM line width is increased:  $0.1\kms$ (\emph{solid}), $0.5\kms$ (\emph{long-dashed}), $0.6\kms$ (\emph{short-dashed}), $0.7\kms$ (\emph{dotted}).  The upper and lower panels show the results obtained neglecting or including local line overlap, respectively (see text). \label{fig:overlap}}
\end{center}
\end{figure}

Line overlap is at present treated \emph{locally} in the OH excitation code, i.e.\ the OH-bearing cloud is treated as a uniform slab with a  velocity width that is everywhere the same.  Its effects are illustrated in Fig.\ \ref{fig:overlap}.  The upper panel shows the maser optical depth as a function of OH column density in a uniform slab when overlap is neglected.  In this case,  line optical depths (ie.~photon escape probabilities) scale as $N_\mathrm{OH}/\Delta v$ so the curves simply shift to the right as $\Delta v$ is increased.  The right hand column shows the effect of including local line overlap: once the line width exceeds $0.5\,\kms$, line overlap starts to shuffle the populations within rotational levels and the inversions are suppressed.

Observed 1720\,MHz maser lines have $\Delta v\sim 1\kms$, at which point local line overlap would suppress the masers, so this treatment overestimates the effect of line overlap.  In reality the OH column does not have a single, uniform, velocity profile throughout -- instead, the column of OH has local line widths that are much smaller than the net velocity width created by the systematic velocity gradient along the line of sight.  This situation requires a ``non-local'' treatment of line overlap effects, in which photons emitted by an infrared transition are absorbed in a neighbouring transition at a different layer in the slab (e.g.~Doel, Gray \& Field 1990).  This form of overlap is less effective, potentially enabling the 1720\,MHz line to remain inverted for larger ($\sim1\kms$) line widths; the 6049\,MHz line however, may then only be weakly inverted at column densities $\sim 10^{17}\ut cm -2 $.  Resolution of this will have to await calculations including non-local overlap.

\section{Discussion}
The lack of detected OH(6049\,MHz) masers in the SNR searches conducted to date may be attributed to a combination if two factors.   First, the OH column densities inferred from GBT observations may be systematically overestimated because of the tendency for OH absorption to overlay regions where the background continuum is brightest.  If the continuum has structure within the beam the optical depth (and so the OH column) will tend to be overestimated, as appears to be the case in 3C\,391 and W44.   However, IC\,443 clump G appears to have a column density $3\ee 17 \ut cm -2 $, more than twice that estimated from the single-dish observations.  Therefore several additional remnants listed in Table~1 may still have OH column densities exceeding the $\sim 10^{17}\ut cm -2 $ threshold suggested for 6049\,MHz inversion.  Second, non-local line overlap might suppress 6049\,MHz inversion more effectively than the 1720\,MHz transition, or simply push the threshold OH column density required for significant maser amplification to columns exceeding $3\ee 17 \ut cm -2 $. 
\acknowledgements
We thank Jack Hewitt for providing his 1720 MHz VLA data.

\newcommand\refitem{\bibitem[]{}}
\newcommand{\ApJ}{\textit{ApJ}}
\newcommand{\apj}{\textit{ApJ}}
\newcommand{\AJ}{\textit{AJ}}
\newcommand{\MNRAS}{\textit{MNRAS}}
\newcommand{\mnras}{\textit{MNRAS}}
\newcommand{\ApJL}{\textit{ApJ} (Letters)}
\newcommand{\apjl}{\textit{ApJ} (Letters)}
\newcommand{\PASA}{\textit{PASA}}
\newcommand{\PASJ}{\textit{PASJ}}
\newcommand{\ApL}{\textit{Astrophys. Lett.}}
\newcommand{\Science}{\textit{Science}}

\end{document}